# Coulomb blockade anisotropic magnetoresistance: Single-electronics meets spintronics

J. Wunderlich*, T. Jungwirth§#, B. Kaestner†, A. C. Irvine¥, A. B. Shick&, K. Wang*, N. Stone*, U. Rana¥, A. D. Giddings#, C. T. Foxon#, R. P. Campion#, D. A. Williams*, B. L Gallagher#

*Hitachi Cambridge Laboratory, Cambridge CB3 0HE, UK

§Institute of Physics ASCR, Cukrovarnická 10, 162 53, Praha 6, Czech Republic

#School of Physics and Astronomy, University of Nottingham, Nottingham NG7 2RD, UK

†National Physical Laboratory, Teddington T11 0LW, UK

¥ University of Cambridge, Microelectronics Research Centre, Cambridge CB3 0HE, UK

&Institute of Physics ASCR, Na Slovance 2, 182 21, Praha 8, Czech Republic

**Single-electronics and spintronics are among the most intensively investigated potential complements or alternatives to CMOS electronics.[1,2,3] Single-electronics, which is based on the discrete charge of the electron[4], is the ultimate in miniaturization and electro-sensitivity and could slash the power consumption of electronic devices. Spintronics, which is based on manipulating electron spins,[5] delivers high magneto-sensitivity and non-volatile memory effects. So far, major developments in the two fields have followed independent paths with only a few experimental studies[6,7] of hybrid single-electronic/spintronic devices. Intriguing new effects have been discovered in such devices but these have not, until now, offered the possibility of useful new functionalities. Here we demonstrate a device which shows a new physical effect, Coulomb blockade anisotropic magnetoresistance, and which offers a route to non-volatile, low-field, and highly electro- and magneto-sensitive operation. Since this new phenomenon reflects the magnetization orientation dependence of the classical single-electron charging energy it does not impose constraints on the operational temperature associated with more subtle quantum effects, such as resonant or spin-coherent tunneling.**

In a single-electron transistor (SET), the transfer of an electron from a source lead to a drain lead via a small, weakly-coupled island is blocked due to the charging energy of $e^2/2C_\Sigma$, where $C_\Sigma$ is the total capacitance of the island.

Applying a voltage $V_G$ between the source lead and a gate electrode changes the electrostatic energy function of the charge $Q$ on the island to $Q^2/2C_\Sigma + QC_GV_G/C_\Sigma$ which has a minimum at $Q_0=-C_GV_G$. By tuning the continuous external variable $Q_0$ to $(n+1/2)e$, the energy associated with increasing the charge $Q$ on the island from $ne$ to $(n+1)e$ vanishes and electrical current can flow between the leads. Changing the gate voltage then leads to Coulomb blockade (CB) oscillations in the source-drain current where each period corresponds to increasing or decreasing the charge state of the island by one electron.

Experiments in SETs in which the leads and island comprise different ferromagnetic materials[6] have shown that in addition to gate voltage dependent CB oscillations it was possible to obtain external magnetic field dependent magneto-CB oscillations. This effect is due to the Zeeman coupling of electron spins to the magnetic field which changes the relative chemical potentials in the leads and in the island, causing a shift in $Q_0$. These magneto-CB oscillations require application of relatively large fields (~ few Tesla) and do not lead to a non-volatile memory effect since the Zeeman shifts in the chemical potential disappear when the magnetic field is turned off. A small low-field hysteretic magnetoresistance (MR) effect has been demonstrated in SETs when the relative orientation of the magnetization of the leads is switched from parallel to antiparallel.[6,7] The sensitivity of the MR to the gate voltage was attributed to resonant tunneling effects through quantized energy levels in the island. In this case the magnetization reorientation causes no shift in $Q_0$ and the effect of the magnetic field is more subtle than that of the gate voltage induced CB oscillations.

Our SETs are fabricated from (Ga,Mn)As, a ferromagnetic semiconductor for which parameters derived from the band structure, including the chemical potential, are strongly anisotropic with respect to the magnetization orientation.[8,9,10] This allows us to induce low-field hysteretic shifts in the CB oscillations, and to demonstrate comparable electro- and magneto-sensitivity and a non-volatile memory effect with "off" state resistances which can be orders of magnitude larger than the "on" state resistances, as we discuss below and in the Supporting Information. A schematic diagram and scanning electron micrographs of the device are shown in Figure 1(a). The SET was fabricated by patterning a side-gated nano-constriction in an ultra-thin (5nm) $Ga_{0.98}Mn_{0.02}As$ epilayer.[10] This technique has been used previously to producing non-magnetic thin film Si and GaAs-based SETs in which disorder potential fluctuations create small islands in the constriction without the need for a lithographically defined island.[11,12] Clear CB conduction diamonds[11] in the $V_G$-$V_{SD}$ plot, where $V_{SD}$ is the source-drain voltage, of one of our (Ga,Mn)As SETs are shown in Figure 1(b). A more detailed discussion of the SET characteristics is given below. We have investigated three different devices all off which show the same qualitative behaviour. Thermal cycling of the individual devices leads to only quantitative changes in the CB oscillation pattern as is typical of SETs realized in narrow constrictions.[11] All the data in this paper are for the SET device of Figure 1.

Figures 1(c) and (d) show the gate-voltage dependent MR of the SET for magnetic fields applied in the plane of the (Ga,Mn)As epilayer and perpendicular to the current direction. The high electro- and magneto-sensitivity and hysteretic behaviour of the device are apparent. For example, the resistance over the constriction decreases by ~100% at 20mT for $V_G$ =0.94 V while it increases by more than 1000% for $V_G$=1.15 V.

We explain below that these resistance changes are associated with (Ga,Mn)As magnetization rotations. The huge and hysteretic MR signals show that our SET can act as a non-volatile memory. In non-magnetic SETs, the CB "on" (low-resistance) and "off" (high-resistance) states are used to represent logical "1" and "0" and the switching between the two states can be realized by applying a gate voltage, in analogy with a standard field-effect transistor. Our ferromagnetic SET can be addressed also magnetically with comparable "on" to "off" resistance ratios in the electric and magnetic modes. The functionality is illustrated in Figure 2. The inset of panel (a) shows two CB oscillation curves corresponding to magnetization states $\mathbf{M_0}$ and $\mathbf{M_1}$. As evident from panel (b), $\mathbf{M_0}$ can be achieved by performing a small loop in the magnetic field, $B \rightarrow B_0 \rightarrow 0$ where $B_0$ is larger than the first switching field $B_{c1}$ and smaller than the second switching field $B_{c2}$, and $\mathbf{M_1}$ is achieved by performing the large field-loop, $B \rightarrow B_1 \rightarrow 0$ where $B_1 < -B_{c2}$. The main plot of Figure 2(a) shows that the high resistance "0" state can be set by either the combinations $(\mathbf{M_1}, V_{G0})$ or $(\mathbf{M_0}, V_{G1})$ and the low resistance "1" state by $(\mathbf{M_1}, V_{G1})$ or $(\mathbf{M_0}, V_{G0})$. We can therefore switch between states "0" and "1" either by changing $V_G$ in a given magnetic state (the electric mode) or by changing the magnetic state at fixed $V_G$ (the magnetic mode). Due to the hysteresis, the magnetic mode represents a non-volatile memory effect.

The diagram in Figure 2(c) illustrates one of the new functionality concepts our device suggests in which low-power electrical manipulation and permanent storage of information are realized in one physical nanoscale element. Panel (d) then highlights the possibility to invert the transistor characteristic; for example, the system is in the low-resistance "1" state at $V_{G1}$ and in the high-resistance "0" state at $V_{G0}$ (reminiscent of an *n*-type field effect transistor) for the magnetization $\mathbf{M_1}$ while the characteristic is inverted (reminiscent of a *p*-type field effect transistor) by changing magnetization to $\mathbf{M_0}$. In Figure S2 of the Supporting Information we also illustrate that in the ferromagnetic semiconductor epilayer used in our study the gate voltage alters significantly the magneto-crystalline anisotropy and can, therefore, be used to trigger magnetization rotations electrically.

The functionality of our hybrid single-electronic/spintronic element arises from a new physical effect, Coulomb blockade anisotropic magnetoresistance (CBAMR). Anisotropic MR is the dependence of the resistance of a ferromagnetic system on the orientation of magnetization relative to the electrical current direction or crystallographic axes. Previous studies of (Ga,Mn)As ferromagnetic semiconductors have shown anisotropic transport characteristics in the ohmic regime (normal AMR) and in the tunneling regime.[8,9] The anisotropic MR of our SET device is demonstrated in Figures 3(b) and (c) and compared with normal AMR in the unstructured part of the (Ga,Mn)As bar (Figure 3(a)). Similar to previous anisotropic MR experiments in (Ga,Mn)As,[9,10,13] a continuous magnetization rotation is achieved by applying a rotating saturation magnetic field (see panel (c) and lower curve in panel (a)) or step-like rotations via intermediate magnetization angles occur in the field sweep measurement (see panel (b) and upper curves in panel (a)) as a result of the combined uniaxial and cubic in-plane anisotropies present in the (Ga,Mn)As epilayer.[14] Here the angle $\theta=0°$ corresponds to $\mathbf{B} \parallel I$, and also to $\mathbf{M} \parallel I$ at saturation fields. In the unstructured part of the bar, higher/lower resistance states correspond to magnetization along/perpendicular to the current direction. Similar behaviour is seen in the SET part of the device at, for example $V_G$=-0.5V, only that the anisotropic MR is now hugely enhanced and depends strongly on the gate voltage.

Figure 4 demonstrates that the dramatic anisotropic MR effect in the SET is due to shifts in the CB oscillation pattern caused by the changes in magnetization orientation. The shifts are clearly seen in the $\theta$-$V_G$ resistance diagram in Figure 4(a) and further highlighted in panel (b), which shows the CB oscillations for several magnetization angles. For example, the oscillations have a peak at $V_G$ = -0.4V for $\theta$=90° which moves to higher gate voltages with increasing $\theta$ and, for $\theta$=130°, the $V_G$= -0.4V state becomes a minimum in the oscillatory resistance. This explains why the magnitude of the resistance variations with $\theta$ at fixed $V_G$ and with $V_G$ at fixed $\theta$ is comparable (see also panels (b) and (c) in Figure 4).

As in the cases of ohmic and tunneling transport[8,9] we attribute the microscopic origin of the CBAMR to anisotropies of a quantity which is derived from the spin-orbit coupled band structure of the ferromagnet. In the CB oscillation regime the anisotropy in the carrier chemical potential is expected to play the dominant role. Schematic diagrams in Figure 4(f) indicate the contributions to the Gibbs energy associated with the transfer of charge $Q$ from the lead to the island. The energy can be written as a sum of the internal, electrostatic charging energy term and the term associated with, in general, different chemical potentials of the lead and of the island: $U=\int_0^Q dQ' \Delta V_D(Q') + Q\Delta\mu/e$, where $\Delta V_D(Q)=(Q+C_GV_G)/C_\Sigma$. The existence of CB effects indicates that the carrier concentration in the (Ga,Mn)As thin film is non-uniform which implies that the difference between chemical potentials of the lead and of the island in the constriction, $\Delta\mu$, will depend on the magnetization orientation. A more detailed theoretical discussion of the sensitivity of the chemical potential anisotropy to the doping density is given in the Supporting Information. Here we refer to the analogy with other parameters derived from the spin-orbit coupled band-structure, such as the tunneling density of states, whose anisotropy is sensitive to the material parameters, typically increasing with decreasing doping in the (Ga,Mn)As.[8,10]

The Gibbs energy $U$ is minimized for $Q_0$=-$C_G(V_G+V_\mathbf{M})$, where the magnetization orientation dependent shift of the CB oscillations is given by $V_\mathbf{M}$=$C_\Sigma/C_G \cdot \Delta\mu(\mathbf{M})/e$. Since $|C_GV_\mathbf{M}|$ has to be of order $|e|$ to cause a marked shift in the oscillation pattern, the corresponding $|\Delta\mu(\mathbf{M})|$ has to be similar to $e^2/C_\Sigma$, i.e., of the order of the island single electron charging energy. From the $I$-$V_C$ characteristics ($V_C$ is the potential drop over the constriction part of the bar) and the temperature dependence of the CB oscillations in our device, we estimate the single electron charging energy to be of the order of few meV (see Figure 4(d) and (e)). This is consistent with the values of $|\Delta\mu(\mathbf{M})|$ obtained by theoretical modelling for typical (Ga,Mn)As epilayer parameters, as shown in Figure S3 in the Supporting Information. We conclude that the CBAMR is based on the classical single-electron charging energy effect and, therefore, the operational temperature is not limited by more subtle quantum effects, such as resonant or spin-coherent tunneling.

Finally we remark that the CBAMR phenomenon and the related single-electronic/spintronic functionalities we observed in (Ga,Mn)As epilayers should be generic to SETs fabricated in ferromagnetic systems with strong spin-orbit coupling. According to our *ab initio* calculations presented in the Supporting Information, chemical potential variations due to magnetization rotations reach 10meV in the FePt ordered alloy.[15] This suggests that, building on the existing expertise in room-temperature non-magnetic single-electronic devices,[16,17] metal based ferromagnetic SETs may offer a route to room-temperature CBAMR applications.

The work was supported by EPSRC, Grant Agency and Ministry of Education of the Czech Republic.

**FIG. 1**

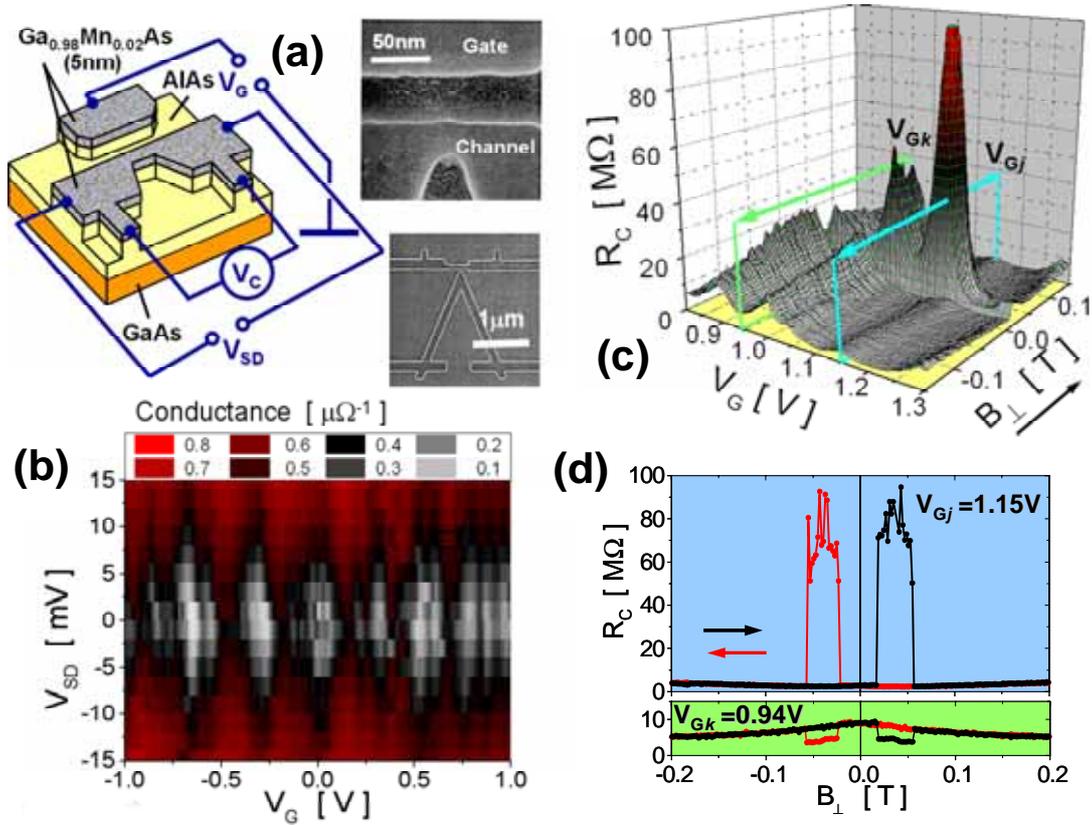

(a) Schematics and scanning electron micrographs of the CBAMR-device. Trench-isolated side-gate and channel aligned along the [110] direction were patterned by e-beam-lithography and reactive ion etching in a 5nm $Ga_{0.98}Mn_{0.02}As$ epilayer grown along the [001] crystal axis on an AlAs buffer layer by low-temperature molecular beam epitaxy. (b) Coulomb-blockade conductance ($I/V_C$) oscillations as a function of gate voltage and source-drain bias at 4.2K. For the detailed SET characterization see Figure 4(c) Resistance ($V_C/I$) measured at $V_{SD}$ = 3mV bias vs. gate voltage and in-plane, perpendicular to the channel direction, magnetic field. The sign and magnitude of the large resistance variations at about 20mT are highly sensitive to the gate-voltage as highlighted in (d) for the gate voltage 1.15V and 0.94V.

**FIG. 2**

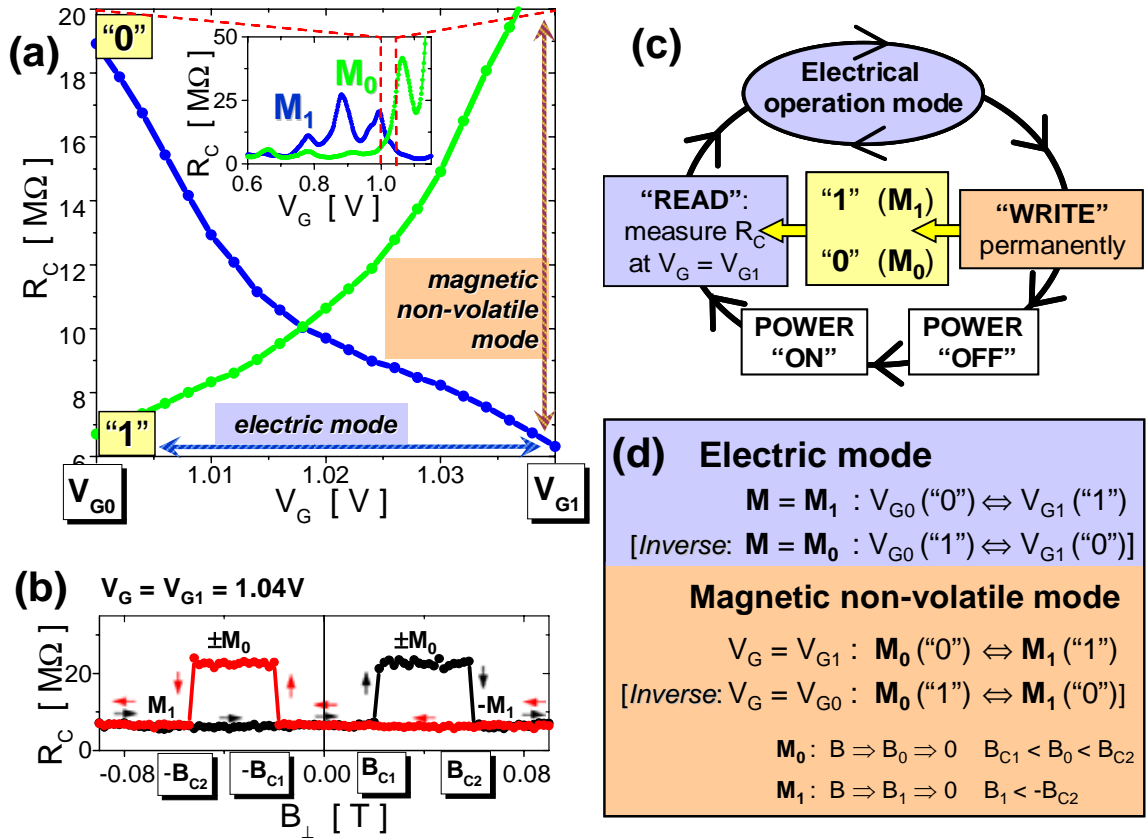

(a) Two opposite transistor characteristics (blue and green) in a gate-voltage range 1V ($V_{G0}$) to 1.04V ($V_{G1}$) for two different magnetization orientations $M_0$ and $M_1$; corresponding Coulomb blockade oscillations in a larger range of $V_G = 0.6$ to 1.15V are shown in the inset. Switching between low-resistance ("1") and high-resistance ("0") states can be performed electrically or magnetically. (b) Hysteretic magnetoresistance at constant gate voltage $V_{G1}$ illustrating the non-volatile memory effect in the magnetic mode. (c) Illustration of integrated transistor (electric mode) and permanent storage (magnetic mode) functions in a single nanoscale element. (d) The transistor characteristic for $M=M_1$ is reminiscent of an *n*-type field effect transistor and is inverted (reminiscent of a *p*-type field effect transistor) for $M=M_0$; the inversion can also be realized in the non-volatile magnetic mode.

**FIG. 3**

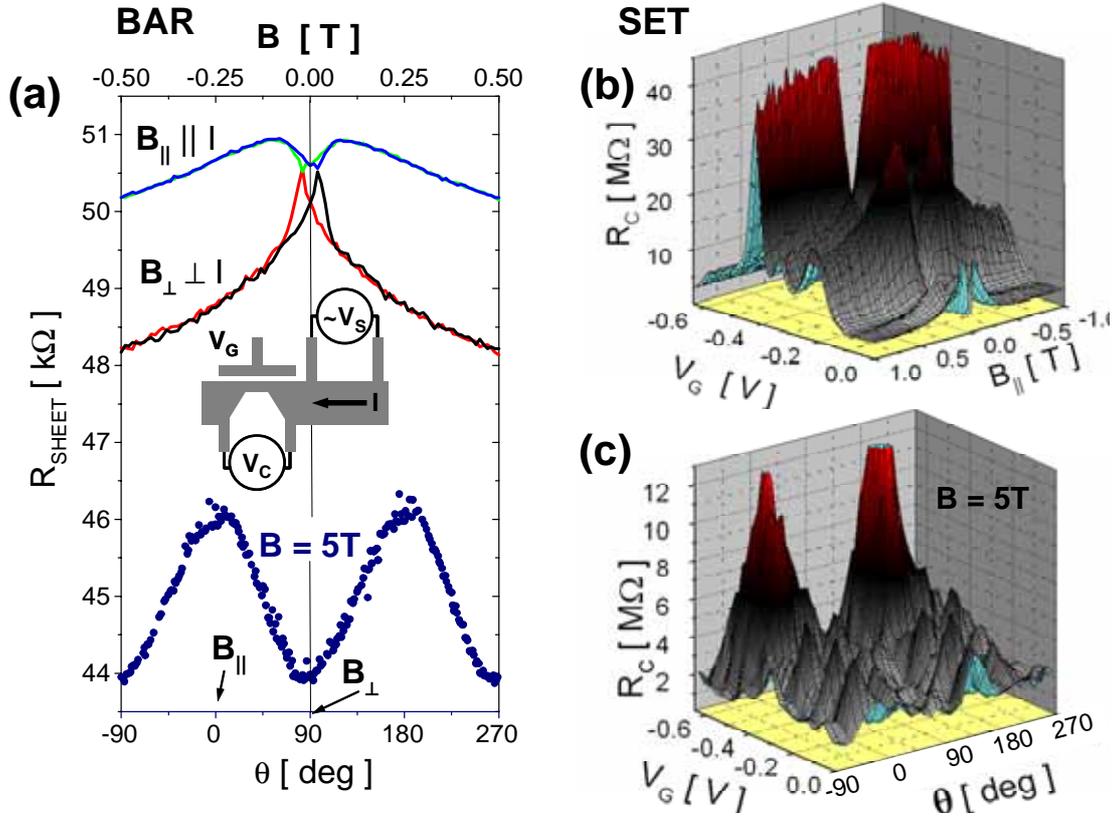

(a) Anisotropic magnetoresistance of the unstructured part of the bar: resistance vs. in-plane magnetic field parallel (blue/green) and perpendicular (red/black) to the current direction. Resistance vs. the angle between the current direction and an applied in-plane magnetic field of constant magnitude of 5T, at which magnetization of the (Ga,Mn)As system is aligned with the external field (bottom curve). (b) and (c) Anisotropic magnetoresistance of the constricted part of the bar: (b) resistance vs. gate voltage and magnetic field applied along the parallel-to-current direction measured at $V_{SD} = 5$mV bias. (c) Resistance vs. gate voltage and the angle between the current direction and an applied in-plane magnetic field of constant magnitude of 5T. Lower overall values of $R_C$ in (c) compared to (b) are due to the isotropic negative MR contribution to the resistance. The same applies to $R_{SHEET}$ in panel (a).

**FIG. 4**

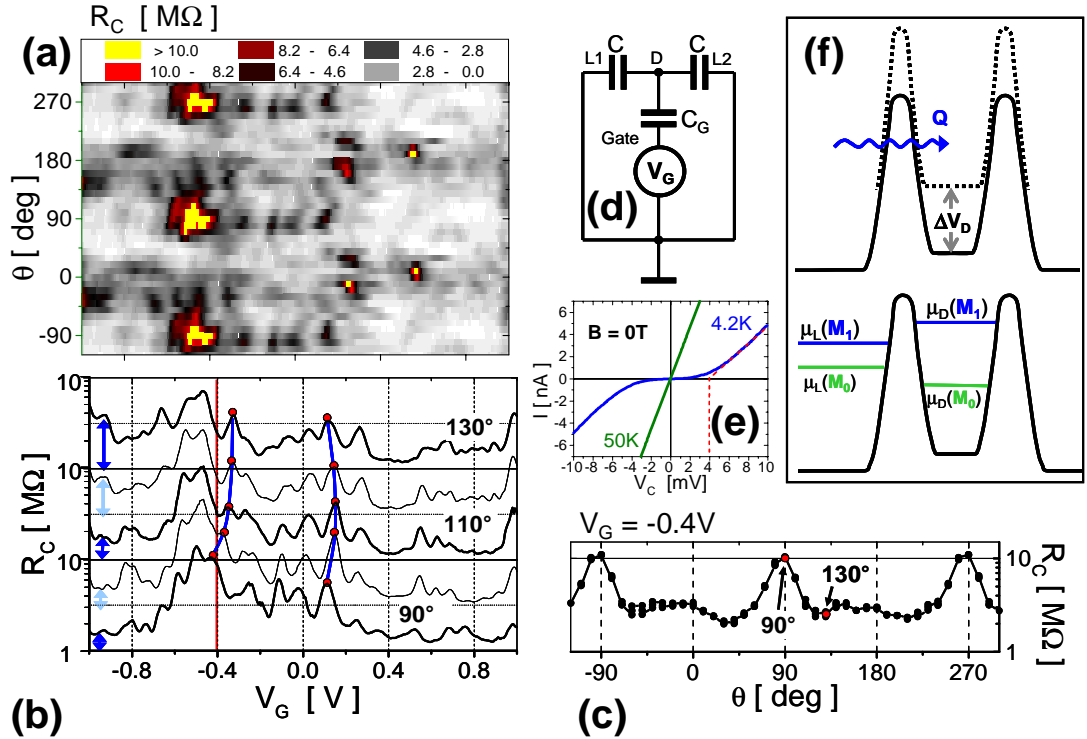

(a) CBAMR in the constricted channel vs. gate voltage and the angle between the current direction and an applied in-plane magnetic field of constant magnitude of 5T measured at $V_{SD}$ = 5mV bias. (b) CBAMR oscillations as a function of the gate voltage for several magnetic field angles at 5T. Blue lines highlight the shifts in the oscillation pattern due to magnetization rotations. The red line shows the development at $V_G$=-0.4V from a local resistance maximum for $\theta$=90° into a local minimum for $\theta$=130°. (c) Resistance as a function of the magnetic field angle for $V_G$=-0.4V. The magnitude of the resistance variations with $\theta$ and with gate voltage is comparable. (d) Circuit model of the SET. The total capacitance $C_\Sigma = 2C + C_G$ defines the Coulomb blockade charging energy $e^2/2C_\Sigma$ which is of the order of a few meV according to the measured data in panel (e); the island-gate capacitance $C_G$ defines the period of the Coulomb blockade oscillations $e/C_G$ which is of the order of ~100mV according to the measured data in panel (b). The inferred lead-island capacitance $C$ is, therefore, approximately 100 times larger than $C_G$. This is due to the large (30nm) gate-constriction, vacuum separation compared to the presumably small lead-island, high dielectric constant (~13) separation. Note that the inferred capacitances represent only approximate characteristic values for our device as more than one island can be expected to be present in the narrow constriction SET. (e) $I$-$V_C$ characteristic of the constriction at 4.2K (Coulomb blockade regime) and 50K (Ohmic regime) used to determine the SET charging energy, as indicated by the red line. This inferred value remains constant with decreasing temperature below 4.2K. (f) Schematics of the charging energy contribution (top sketch) to the Gibbs energy and the contribution proportional to the difference in chemical potentials of the lead and of the island (bottom sketch). The second term depends on magnetization orientation causing the CBAMR effect.

[1] Siegel, R. W., Hu, E., Cox, D. M. et al. Nanostructure science and Technology. *World Technology panel report*, 67-88 (1999). http://www.wtec.org/loyola/nano/toc.htm.

[2] Single electron transistors. *Technology Review, Special Report:R&D '04*, 4 (December 2004). http://www.technologyreview.com/InfoTech/wtr_13988,294,p4.html.

[3] Magnetoresistive random access memories (MRAM) summary. *Research and Markets*, (June 2005). http://www.researchandmarkets.com/reports/c21851.

[4] Likharev, K. K. Single-electron devices and their applications. *Proceedings of the IEEE*, **87**, 606-632 (1999).

[5] Wolf, S. A., Awschalom, D.D., Buhrman, R. A. et al. Spintronics: A Spin-Based Electronics Vision for the Future. *Science* **294**, 1488-1495 (2001).

[6] Ono, K., Shimada, H., and Ootuka,Y. Enhanced magnetic valve effect and magneto-Coulomb oscillations in ferromagnetic single electron transistor. *J. Phys. Soc. Jpn.* **66**, 1261-1264 (1997).

[7] Sahoo, S., Kontos, T., Furer, J., Hoffmann, C., Gräber, M., Cottet, A., and Schönenberger, C. Electric field control of spin transport. *Nature Physics* **1**, 99-102 (2005).

[8] Jungwirth, T., Sinova J., Wang, K. Y. et al. DC-transport properties of ferromagnetic (Ga,Mn)As semiconductors. *Appl. Phys. Lett.* **83**, 320-322 (2003).

[9] Gould, C., Rüster C., Jungwirth, T. et al. Tunneling Anisotropic Magnetoresistance: A spin-valve like tunnel magnetoresistance using a single magnetic layer, *Phys. Rev. Lett.* **93**, 117203 1-4 (2004).

[10] A.D. Giddings, A.D., Khalid, M.N., Jungwirth, T., Wunderlich, J., et al. Large tunneling anisotropic magnetoresistance in (Ga,Mn)As nanoconstriction. *Phys. Rev. Lett.* **94**, 127202 1-4 (2005).

[11] Kastner, M.A. The single-electron transistor. *Rev. Mod. Phys.* **64**, 849-858 (1992).

[12] Tsukagoshi, K., Alphenaar, B. W., and Nakazato, K. Operation of logic function in a Coulomb blocade device. *Appl. Phys. Lett.* **73**, 2515-2517 (1998).

[13] Rüster, C., Gould, C., Jungwirth, T., Sinova, J., Schott, G. M., Giraud, R., Brunner, K., Schmidt, G., and Molenkamp, L. W. Very Large Tunneling Anisotropic Magnetoresistance of a (Ga,Mn)As/GaAs/(Ga,Mn)As Stack, *Phys. Rev. Lett.* **94**, 027203 1-4 (2005).

[14] Sawicki, M., Wang, K-Y., Edmonds, K. W., Campion, R. P., Staddon, C. R., Farley, N. R. S., Foxon, C. T., Papis, E., Kaminska, E., Piotrowska, A., Dietl, T., Gallagher, B.L. In-plane uniaxial anisotropy rotations in (Ga,Mn)As thin films. *Phys. Rev. B* **71**, 121302 (2005).

[15] Shick, A. B. and Mryasov, O. N. Coulomb correlations and magnetic anisotropy in ordered $L1_0$ CoPt and FePt alloys. *Phys. Rev. B* **67**, 172407 1-4 (2003).

[16] Matsumoto, K., Ishii, M., Segawa, K., Oka, Y., Vartanian, B. J., Harris, J. S. Room temperature operation of a single electron transistor made by the scanning tunneling microscope nanooxidation process for the $TiO_x$/Ti system. *Appl. Phys. Lett.* **68**, 34-36 (1996).

[17] Saitoh, M., Harata, H., Hiramoto, T. Room-temperature demonstration of low-voltage and tunable static memory based on negative differential conductance in silicon single-electron transistors. *Appl. Phys. Lett.* **85**, 6233-6235 (2004).

**Supporting information for the article "Coulomb blockade anisotropic magnetoresistance: Single-electronics meets spintronics"**

**Large Magnetoresistance**

Very large magneto-resistance variation due to the CBAMR effect where also found when the magnetization was rotated out of the plane of the (Ga,Mn)As epilayer by either applying a large rotating magnetic field or during the magnetic field sweep experiment with the field aligned perpendicular to the epilayer plane. As shown in Figure S1, the resistance changes by more than 3 orders of magnitude at certain gate voltages.

**Gate voltage dependent magnetization switching**

In the main text of the article we demonstrated that small changes in the gate voltage result in CB resistance oscillations and that these oscillations are shifted when magnetization is rotated. Here we illustrate that in the ferromagnetic semiconductor epilayer used in our study, larger gate voltage changes can also alter significantly the magneto-crystalline anisotropy profile and can, therefore, be used to trigger magnetization rotations electrically.[1] In Figure S2 the strong gate voltage dependence of a switching field is highlighted by the red dashed line in panel (a). The experimental demonstration of electrical field assisted magnetization reorientation is shown in panel (b). Note that variations of the uniaxial and cubic anisotropies with carrier doping have already been reported in (Ga,Mn)As epilayers.[2] Depending on the carrier concentration, either uniaxial or biaxial (cubic) anisotropy were dominant. In our device, the carrier density may vary with the gate voltage in the electric field exposed regions.

**Theoretical modelling of chemical potential anisotropies**

The calculation of spectral properties of (Ga,Mn)As ferromagnetic semiconductors using *ab initio* approaches is challenging due, partly, to a strongly correlated nature of Mn local moments and their random distribution in the lattice. Instead, the effective kinetic-exchange model have been frequently used to study the properties of these dilute magnetic moment *p*-type semiconductors, particularly the properties related to the strong spin-orbit coupling in the valence band.[3,4,5,6] Here we use the model to illustrate the changes in the chemical potential caused by magnetization rotations. The calculations are done for zero temperature. The model assumes a cubic magnetocrystalline anisotropy associated with the zinc-blende crystal structure of the semiconductor. An additional uniaxial field observed in experiments is modeled[2] by introducing a weak shear strain, $e_{xy}$=0.001. Since the calculations neglect size quantization effects in the ultra-thin (Ga,Mn)As epilayer and any asymmetry of the spin-splitting of the valence band with respect to the band edge of the pure GaAs, the results should be regarded as only qualitative estimates. The data are presented in the main panel of Figure S3 for Mn concentrations 2% and 5% and over a range of carrier concentrations where the higher values are expected to correspond to the leads and the lower values to the constricted part of the bar. The data confirm that the chemical potential anisotropies are sensitive to the variations in the density of carriers and local moments (may even change sign), and that |$\Delta\mu$| of order ~1meV inferred from the experiment are plausible.

In ferromagnetic metals, highly accurate *ab initio* methods are available to study the spin-orbit coupled spectral properties. Recently, the technique has been successfully applied to describe magneto-crystalline anisotropies in L1$_0$ FePt and CoPt ordered alloys and to predict their tunneling density of states anisotropies.[7,8] In these systems, the transition metal produces large exchange splitting resulting in the Curie temperatures well above room temperature while the heavy elements of Pt substantially increase the strength of spin-orbit coupling in the band structure of the alloy. We use the systems as an illustration of the prospect for high temperature CBAMR in ferromagnetic metals. The calculated chemical potential anisotropies, shown in the inset of Figure S3, are roughly an order of magnitude larger than in (Ga,Mn)As. As discussed in the main text of the article, a similar enhancement may be expected for the corresponding temperature limit of the CBAMR operation.

FIG. S1

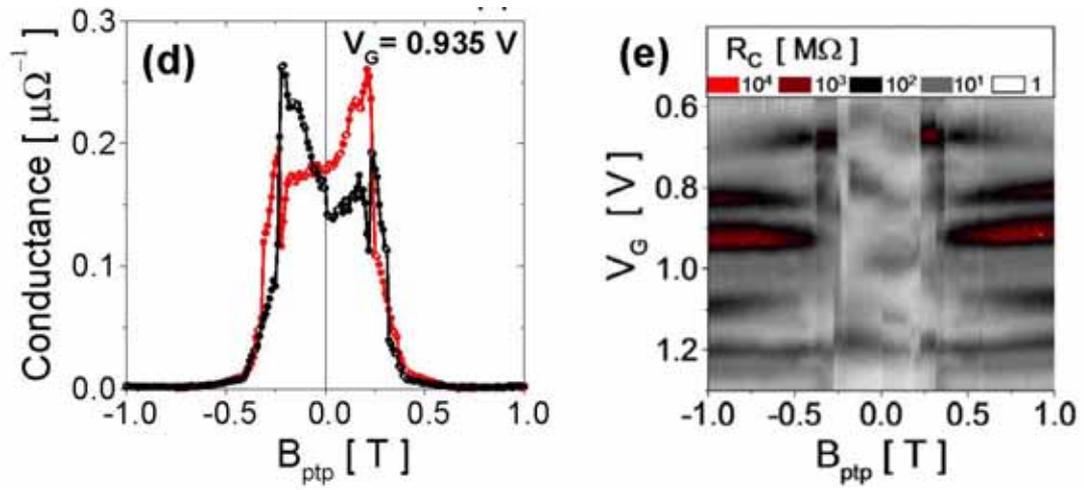

(a) Resistance vs. gate-bias $V_G$ and perpendicular-to-plane oriented magnetic field $B_{ptp}$ measured at $V_{SD}$ = 5mV. (b) Very large conductance vs. $B_{ptp}$ variation at the gate-voltage of $V_G$ = 0.935V. The system becomes practically insulating at saturation.

FIG. S2

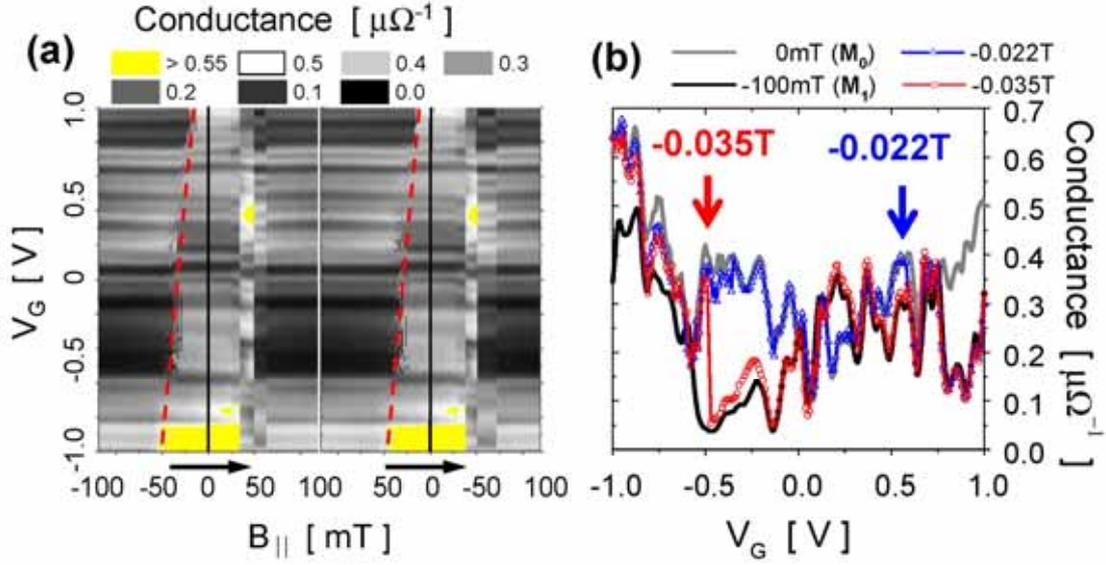

(a) Conductance measurements at $V_{SD} = 3$mV bias vs. gate voltage and in-plane, parallel to channel magnetic field. The dashed red line highlights the critical reorientation field which is strongly gate bias dependent and decreases from about 0.04T at $V_G = -1$V down to less than 0.02T at $V_G = 1$V. (b) CB oscillations at zero magnetic field where the system remains at magnetization $M_0$ (grey) and at field -0.1T where the system remains at magnetization $M_1$ (black) over the gate voltage range between $V_G = -1$V to 1V. Conductance measurements at intermediate field strengths of -0.022T (blue) and -0.035T (red) show a transition from $M_0$ to $M_1$ at critical gate voltages of about 0.6V and -0.5V, respectively.

FIG. S3

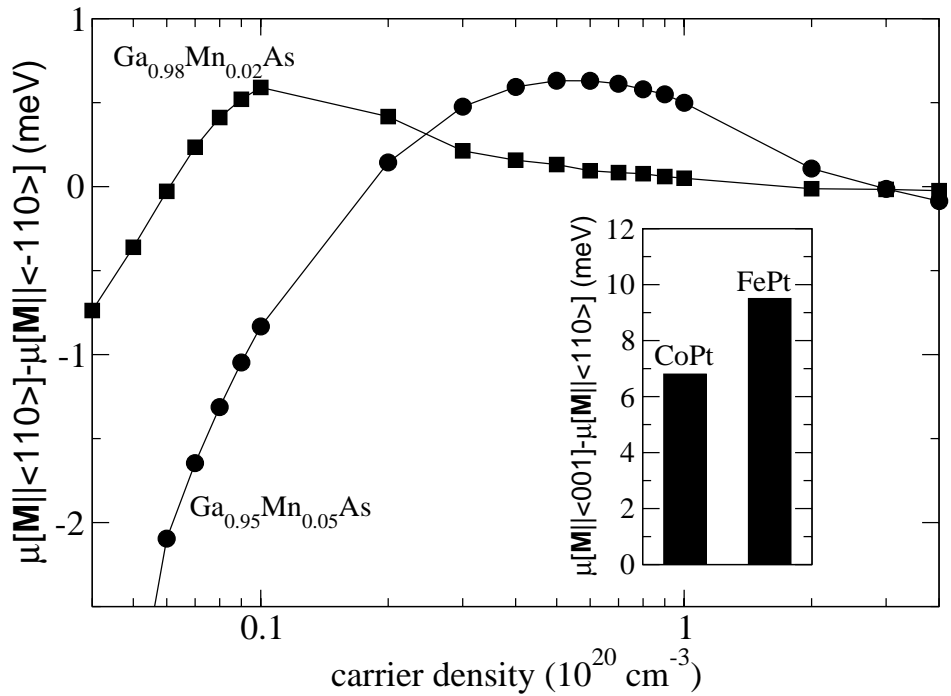

Main plot: kinetic-exchange model calculations of the chemical potential anisotropy in (Ga,Mn)As as a function of carrier dansity for Mn local moment concentrations of 2% and 5%. The uniaxial anisotropy is modeled by introducing a weak shear strain $e_{xy}$=0.001. Inset: *ab initio* calculations of chemical potential anisotropies in $L1_0$ FePt and CoPt ordered alloys.

[1] Chiba, D., Yamanouchi, M., Matsukura, F., and Ohno, H. Electrical Manipulation of Magnetization Reversal in a Ferromagnetic Semiconductor. *Science* **301**, 943 (2003).

[2] Sawicki, M., Wang, K-Y., Edmonds, K. W., Campion, R. P., Staddon, C. R., Farley, N. R. S., Foxon, C. T., Papis, E., Kaminska, E., Piotrowska, A., Dietl, T., Gallagher, B.L. In-plane uniaxial anisotropy rotations in (Ga,Mn)As thin films. *Phys. Rev. B* **71**, 121302 (2005).

[3] Dietl, T., Ohno, H., Matsukura, F., Cibert, J., Ferrand, D. Zener model description of ferromagnetism in zinc-blende magnetic semiconductors. *Science* **287**, 1019 (2000).

[4] Abolfath, M., Jungwirth, T., Brum, J., and MacDonald, A.H. Theory of magnetic anisotropy in $III_{1-x}Mn_xV$ Ferromagnets. *Phys. Rev. B* **63**, 054418 (2001).

[5] Jungwirth, T., Sinova J., Wang, K. Y. et al. DC-transport properties of ferromagnetic (Ga,Mn)As semiconductors. *Appl. Phys. Lett.* **83**, 320-322 (2003).

[6] Gould, C., Rüster C., Jungwirth, T. et al. Tunneling Anisotropic Magnetoresistance: A spin-valve like tunnel magnetoresistance using a single magnetic layer. *Phys. Rev. Lett.* **93**, 117203 1-4 (2004).

[7] Shick, A. B. and Mryasov, O. N. Coulomb correlations and magnetic anisotropy in ordered $L1_0$ CoPt and FePt alloys. *Phys. Rev. B* **67**, 172407 1-4 (2003).

[8] Shick, A. B., Máca, F., Mašek, J., Jungwirth, T. Prospect for room temperature tunneling anisotropic magnetoresistance effect: Density of states anisotropies in CoPt systems. *Phys. Rev. B* in press.